\documentclass[
    ,final            
  ]
  {aipproc}

\layoutstyle{6x9}

\usepackage{graphicx}
\usepackage{amsmath}
\usepackage{color}

\definecolor{darkgreen}{rgb}{0,0.5,0}
\definecolor{purple}{rgb}{0.5,0,0.5}
\definecolor{nblue}{rgb}{0.0,0.0,0.50}
\definecolor{scarlet}{rgb}{1.0,0.2,0}

\newcommand{\lsim}{\mathrel{\rlap{\lower4pt\hbox{\hskip0pt$\sim$}}
\raise1pt\hbox{$<$}}}           
\newcommand{\gsim}{\mathrel{\rlap{\lower4pt\hbox{\hskip0pt$\sim$}}
\raise1pt\hbox{$>$}}}           

\begin{document}

\title{Opportunities and Challenges for Theory \\ in the N$^\ast$ program}

\classification{%
12.38.Aw, 	
12.38.Lg, 	
13.40.-f, 	
14.20.Gk	
}
\keywords      {Baryon and meson spectra, Baryon transition form factors,  Confinement, Dynamical chiral symmetry breaking}

\author{C.\,D.~Roberts$^{1,2,3,4}$}
{address={
  $^1$Physics Division, Argonne National Laboratory, Argonne, Illinois 60439, USA\\
  $^2$Institut f\"ur Kernphysik, Forschungszentrum J\"ulich, D-52425 J\"ulich, Germany\\
  $^3$Department of Physics, Illinois Institute of Technology, Chicago, Illinois 60616-3793, USA\\
  $^4$Department of Physics, Center for High Energy Physics and State Key Laboratory of Nuclear Physics and Technology, Peking University, Beijing 100871, China
  }}

\begin{abstract}
The $N^\ast$-program provides a path to understanding the essentially-nonperturbative fundamentals at the heart of the Standard Model: confinement and dynamical chiral symmetry breaking.  Relating this data to QCD's basic degrees-of-freedom is a key challenge for theory.  In tackling it, one steps immediately into the domain of relativistic quantum field theory where within the key phenomena can only be understood via nonperturbative methods.  No one tool is yet fully equal to the challenge.  Nonetheless, the last few years have seen significant progress in QCD-based theory, and the reaction models necessary to bridge the gap between that theory and experiment.
\end{abstract}

\maketitle

\hspace*{-\parindent}\mbox{\textbf{1.~Introduction}}~~
Two key questions drive the international effort focused on the physics of excited nucleons:
Which hadron states and resonances are produced by QCD, the strongly-interacting part of the Standard Model, and how are they constituted?  In addressing these questions, the $N^\ast$ program stands alongside the search for hybrid and exotic mesons as an integral part of the search for an understanding of quantum chromodynamics (QCD).
Elucidating the real-world predictions that follow from QCD is basic to drawing the map that explains how the Universe is constructed.  With QCD, Nature has defined the sole known example of a strongly-interacting quantum field theory that is defined by degrees-of-freedom which cannot directly be detected.  This empirical fact of \emph{confinement} ensures that QCD is the most interesting and challenging piece of the Standard Model.  Moreover, there are many reasons to believe that QCD generates forces so strong that less-than 2\% of a nucleon's mass can be attributed to the so-called current-quark masses that appear in QCD's Lagrangian; viz., forces capable of generating mass from nothing, a phenomenon known as dynamical chiral symmetry breaking (DCSB).

Neither confinement nor DCSB is apparent in QCD's Lagrangian.  Yet they play the dominant role in determining the observable characteristics of real-world QCD.  The physics of hadrons is ruled by such \emph{emergent phenomena}, which can only be explained through the use of nonperturbative methods in quantum field theory.  This is both the greatest novelty and the greatest challenge within the Standard Model.  We must find essentially new ways and means to explain precisely the observable content of QCD.  Building a bridge between QCD and the observed properties of hadrons is one of the key problems for modern science.

\begin{figure}[t]
\includegraphics[clip,width=0.45\textwidth]{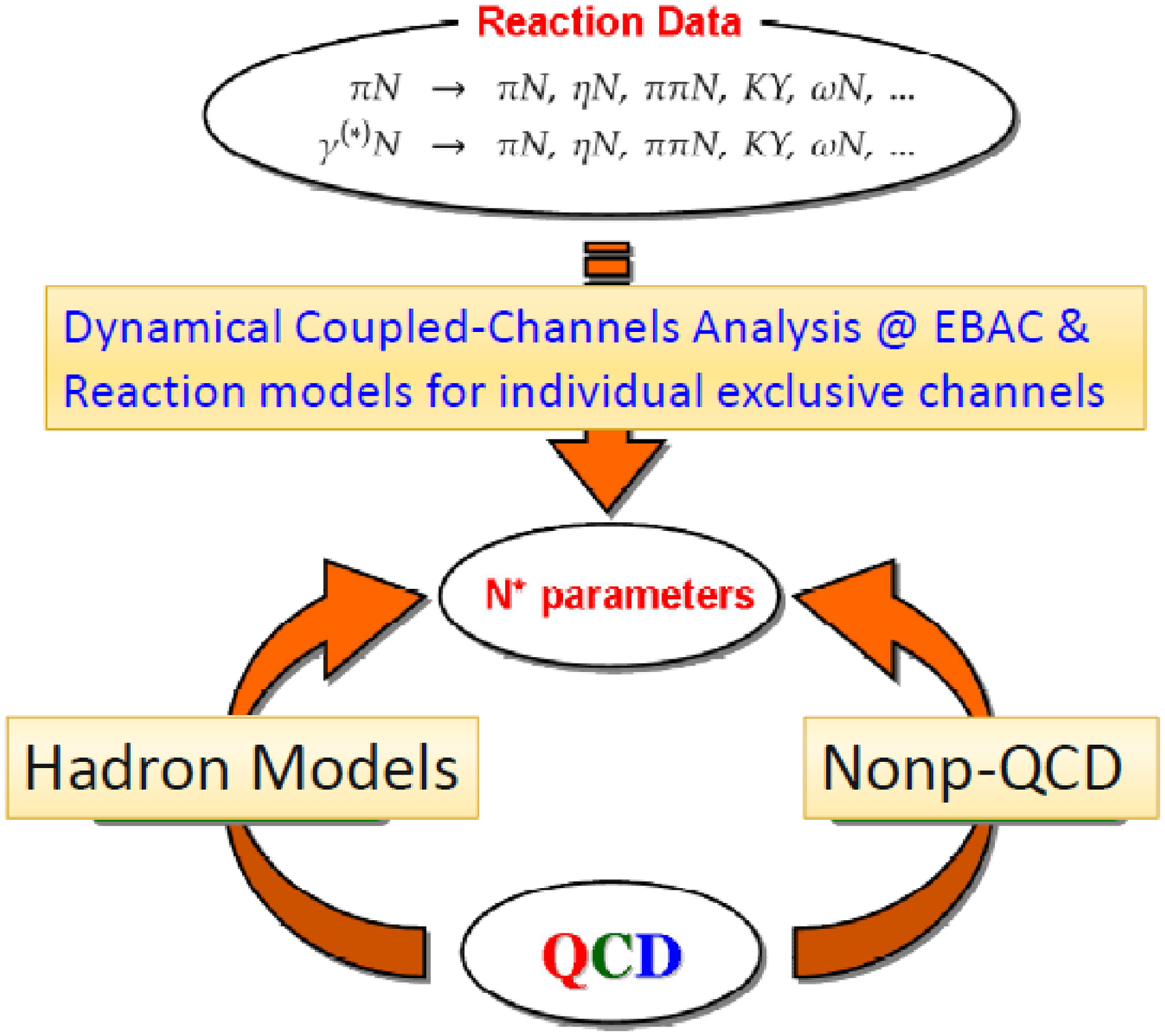}\hspace*{2em}



\parbox{0.45\textwidth}{$\,$\vspace*{-32ex}\\
{\footnotesize \textbf{\footnotesize FIGURE 1}.~~Flow chart depicting the strategy of the $N^\ast$ program.  (Adapted from Ref.\,\protect\cite{Burkert:2007zza}.)}}

\end{figure}
\addtocounter{figure}{1}

The operating model of the $N^\ast$-program is illustrated in Fig.\,1.   Given a large body of data from exclusive meson photo-, electro- and hadro-production off nucleons, a well-constrained reaction theory must be employed to extract reliable information on hadron resonance parameters and transition form factors.  This information provides the target for theoretical tools based on QCD, which must relate it to the nonperturbative, strong-interaction mechanisms that are responsible for hadron structure and resonance formation.  In this way one may reach an understanding of how the interactions between dressed-quarks and -gluons create ground and excited nucleon states; and how these interactions emerge from QCD.  Material progress has been achieved in this way since the last meeting \cite{NStar2009Proc}, some part of which is described in Ref.\,\cite{Aznauryan:2011zz}.

Numerous theoretical tools have been deployed within the $N^\ast$-program.  A sophisticated reaction theory has been developed and applied, with the results subjected to the scrutiny of QCD-based tools, amongst them: constituent-quark- and algebraic-models; Dyson-Schwinger equations (DSEs); generalized parton distributions; lattice-regularized QCD; and light-cone sum rules.  This proceedings volume provides a snapshot of recent progress in these areas.

\bigskip

\hspace*{-\parindent}\mbox{\textbf{2.~Confinement}}~~A driving force behind the $N^\ast$-program is the need to understand confinement.  However, regarding the nature of confinement, little is known and much is misapprehended.   It is thus important to state that the potential between infinitely-heavy quarks measured in simulations of quenched lattice-QCD -- the so-called static potential -- is \emph{irrelevant} to confinement in the real world, in which light quarks are ubiquitous.  It is a basic feature of QCD that light-particle creation and annihilation effects are essentially nonperturbative.  It is therefore impossible in principle to compute a potential between two light quarks \cite{Bali:2005fu,Chang:2009ae}.  Hence, in discussing this physics, linearly rising potentials, flux-tube models, etc., have no connection nor justification within QCD.

\begin{figure}[t]
\includegraphics[clip,width=0.70\textwidth]{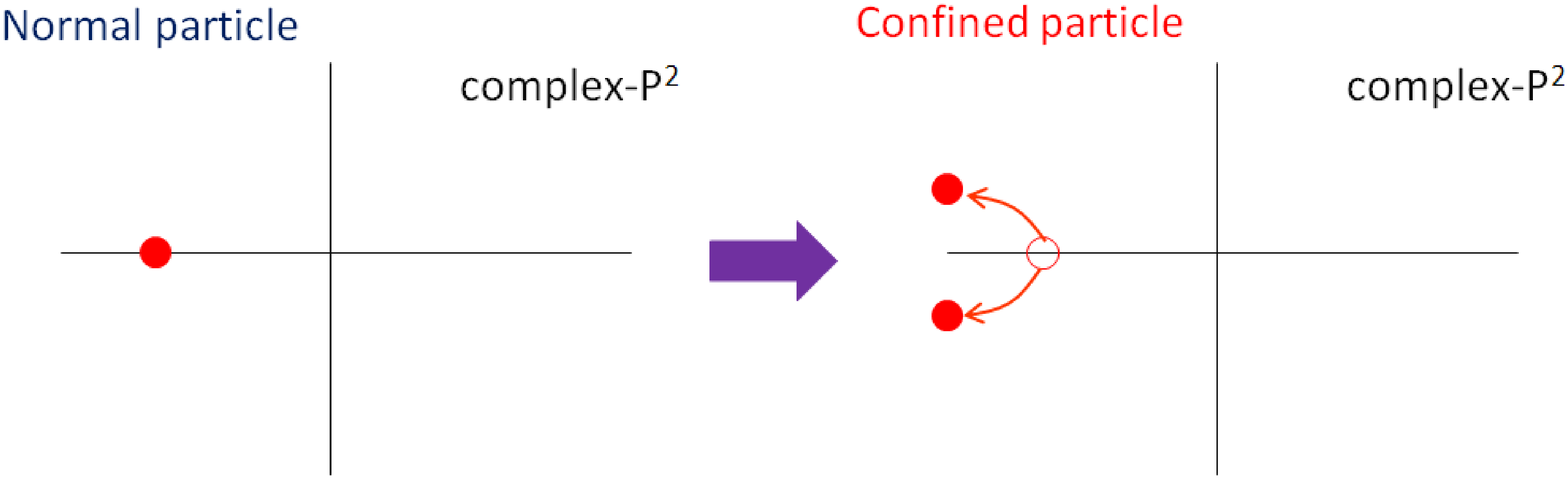}
\caption{\label{fig:confined} \emph{Left panel} -- An observable particle is associated with a pole at timelike-$P^2$.  This becomes a branch point if, e.g., the particle is dressed by photons.  \emph{Right panel} -- When the dressing interaction is confining, the real-axis mass-pole splits, moving into pairs of complex conjugate poles or branch points.  No mass-shell can be associated with a particle whose propagator exhibits such singularity structure.}
\end{figure}

\begin{figure}[t]

\centerline{
\includegraphics[clip,width=0.48\textwidth]{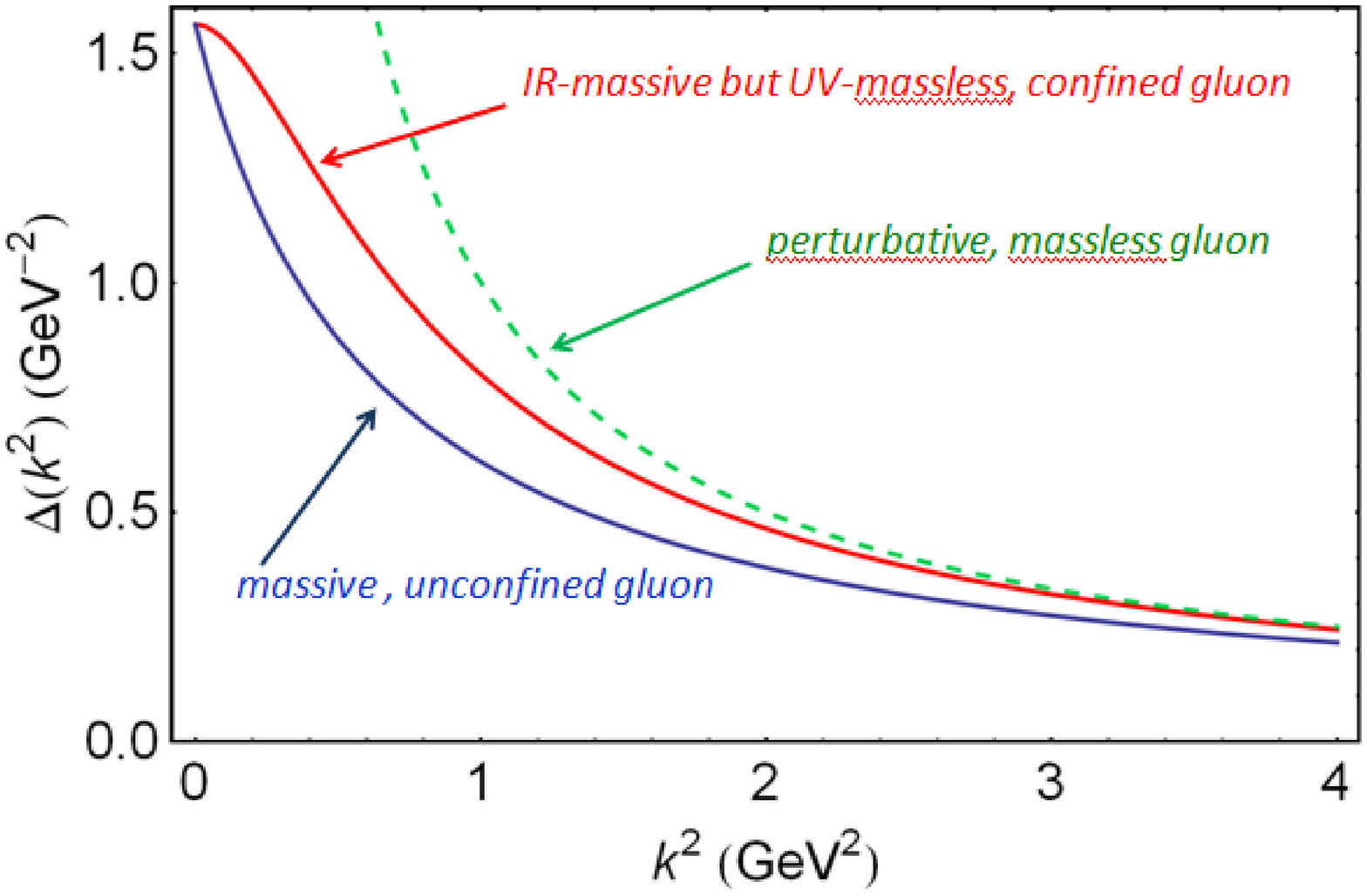}}\hspace*{-8em}
\includegraphics[clip,width=0.48\textwidth]{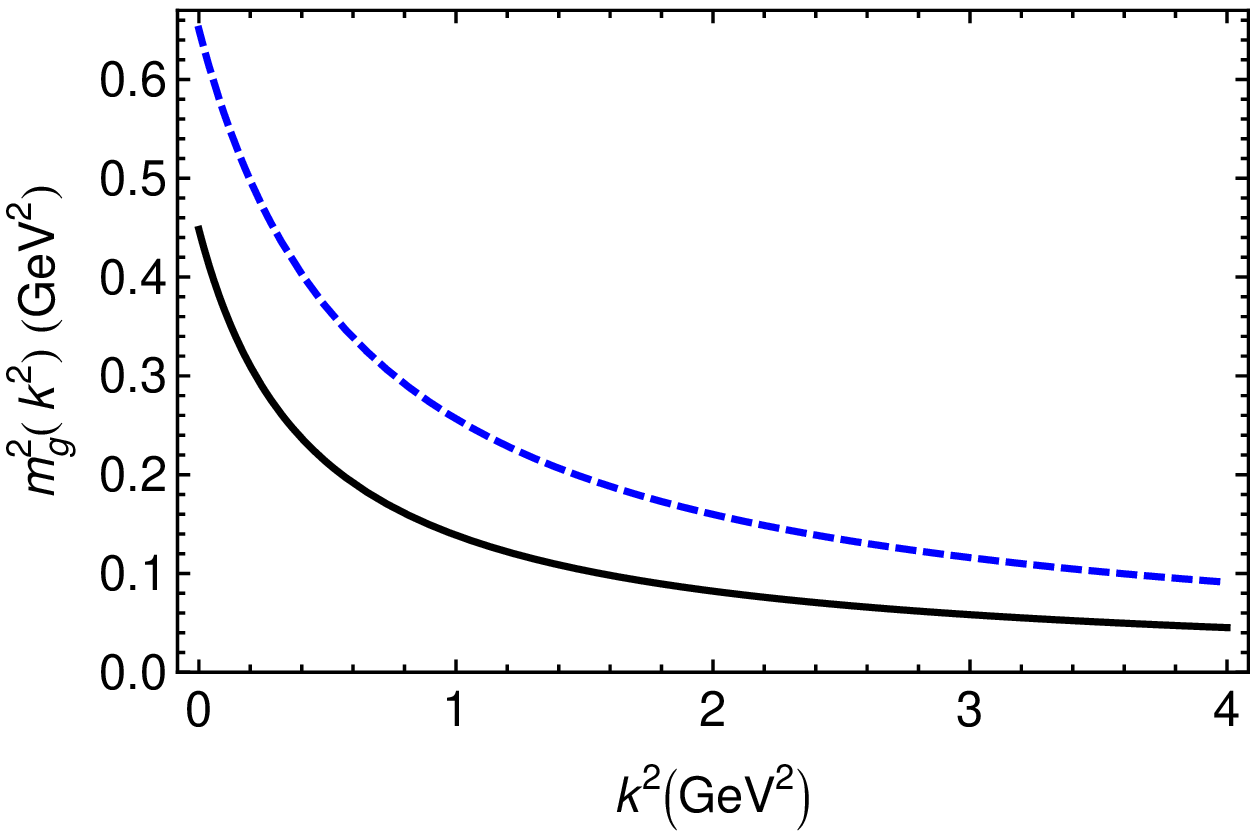}\hspace*{10em}

\caption{\label{fig:gluonrp}
\emph{Left panel} --
$\Delta(k^2)$, the function that describes dressing of a gluon propagator, plotted for three distinct cases.
A bare gluon is described by $\Delta(k^2) = 1/k^2$ (dashed line), which is plainly convex on $k^2\in (0,\infty)$.  Such a propagator is associated with an observable particle.
In some theories, interactions generate a mass in the transverse part of the gauge-boson propagator, so that $\Delta(k^2) = 1/(k^2+m_g^2)$.  This is nevertheless a convex function.
In QCD, however, self-interactions generate a momentum-dependent mass for the gluon, which is large at infrared momenta but vanishes in the ultraviolet \protect\cite{pepe:11}.  This is illustrated by the curve labeled ``IR-massive but UV-massless.''  With the generation of a mass-\emph{function}, $\Delta(k^2)$ exhibits an inflexion point and hence the gluon cannot propagate to a detector.
\emph{Right panel} -- The most successful symmetry-preserving studies of $\pi$, $\rho$ and nucleon properties indicate that the gluon mass function has the form illustrated here.  They are distinguished by their infrared value: solid-curve, $M_g=0.67\,$GeV; and dashed-curve, $M_g=0.81\,$GeV.  At present, both are acceptable but future developments in the $N^\ast$-program should assist in distinguishing between them.
(Adapted from Refs.\,\protect\cite{Chang:2011vu,Qin:2011}.)
}
\end{figure}

On the other hand, confinement can be related to the analytic properties of QCD's Schwinger functions; i.e., the dressed-propagators and -vertices.  
This perspective was laid out in Ref.\,\cite{Krein:1990sf}.  Whilst there is a great deal of mathematical background to this observation, it is readily illustrated, Fig.\,\ref{fig:confined}.  The simple pole of an observable particle produces a propagator that is a monotonically decreasing convex function, whereas the evolution depicted in the right-panel of Fig.\,\ref{fig:confined} is manifest in the propagator as the appearance of an inflexion point at $P^2 > 0$.  To complete the illustration, consider $\Delta(k^2)$, which is the single scalar function that describes the dressing of a Landau-gauge gluon propagator.  Three possibilities are exposed in Fig.\,\ref{fig:gluonrp}.  The inflexion point possessed by $M(p^2)$, visible in Fig.\,4, entails, too, that the dressed-quark is confined.

From the perspective that confinement can be related to the analytic properties of QCD's Schwinger functions, the question of light-quark confinement may be translated into the challenge of charting the infrared behavior of QCD's \emph{universal} $\beta$-function. (Of course, the behavior of the $\beta$-function on the perturbative domain is well known.)  This is a well-posed problem whose solution is an elemental goal of modern hadron physics; e.g., Refs.\,\cite{Qin:2011,Brodsky:2010ur,Aguilar:2010gm}.  It is the $\beta$-function that is responsible for the behavior evident in Figs.\,\ref{fig:gluonrp} and 4, and thereby the scale-dependence of the structure and interactions of dressed-gluons and -quarks.  One of the more interesting of contemporary questions is whether it is possible to reconstruct the $\beta$-function, or at least constrain it tightly, given empirical information on the gluon and quark mass functions.  Experiment-theory feedback within the $N^\ast$-program shows promise for providing the latter.

\begin{figure}[t]

\centerline{
\hspace*{-16em}\includegraphics[clip,width=0.50\textwidth]{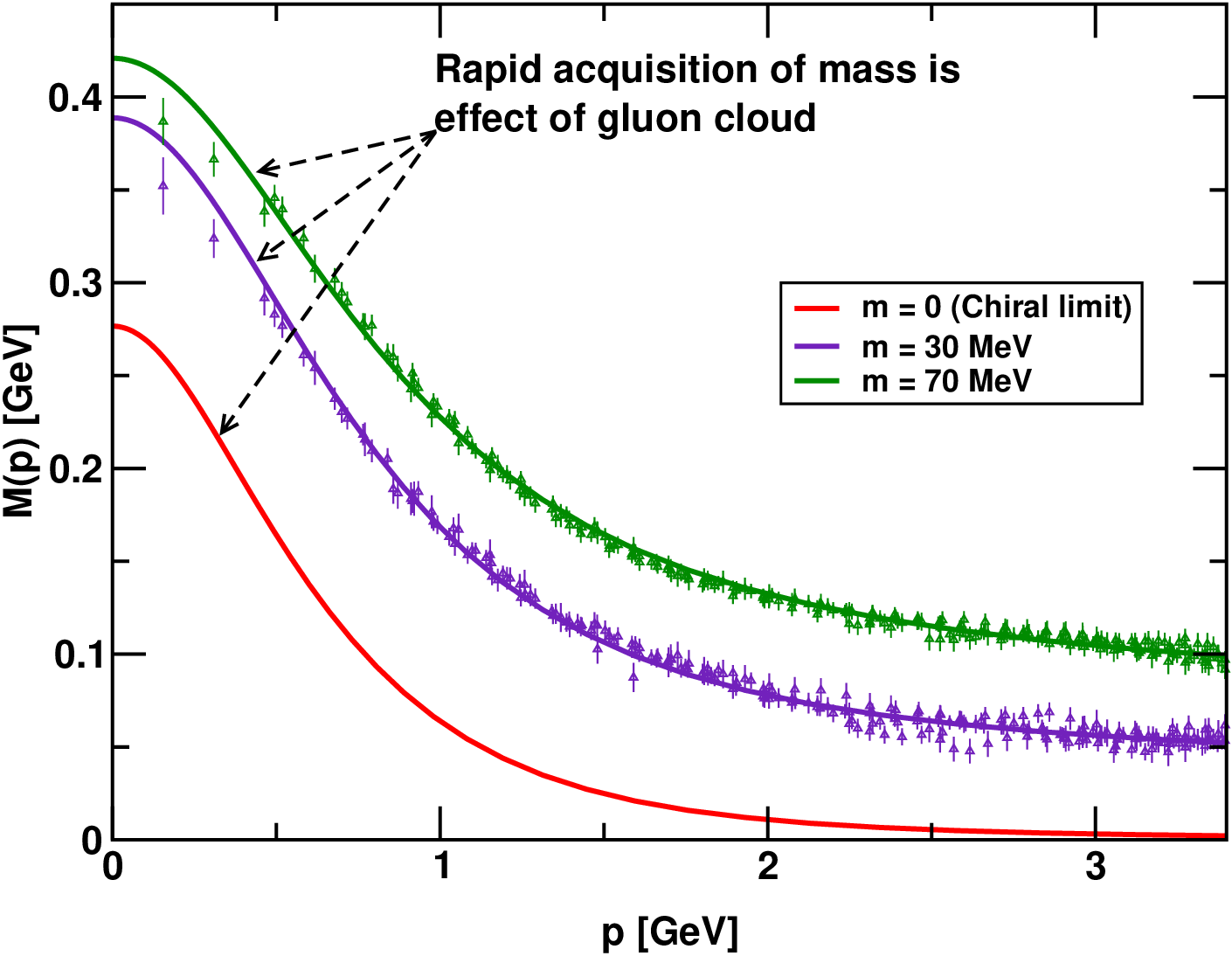}}\hspace*{-15em}

\parbox{0.40\textwidth}{$\,$\vspace*{-32ex}\\
{\footnotesize \textbf{\footnotesize FIGURE 4}.~~Dressed-quark mass function, $M(p)$: \emph{solid curves} -- DSE results, \protect\cite{Bhagwat:2006tu}, ``data'' -- lattice-QCD simulations \protect\cite{Bowman:2005vx}.  (NB.\ $m=70\,$MeV is the uppermost curve.  Current-quark mass decreases from top to bottom.)  
The constituent mass arises from a cloud of low-momentum gluons attaching themselves to the current-quark: DCSB is a truly nonperturbative effect that generates a quark mass \emph{from nothing}; namely, it occurs even in the chiral limit, as evidenced by the $m=0$ curve.
(Adapted from Ref.\,\protect\cite{Bhagwat:2007vx}.)}}
\end{figure}
\addtocounter{figure}{1}

\bigskip

\hspace*{-\parindent}\mbox{\textbf{3.~Dynamical Chiral Symmetry Breaking}}~~Whilst the nature of confinement \label{`confinement'} is still debated, Fig.\,4 shows that DCSB is a fact.  It is the most important mass generating mechanism for visible matter in the Universe, being responsible for roughly 98\% of the proton's mass \cite{Flambaum:2005kc}.  Indeed, the Higgs mechanism is (almost) irrelevant to light-quarks.  In Fig.\,4 one observes the current-quark of perturbative QCD evolving into a constituent-quark as its momentum becomes smaller.
%
%
This behavior, and that illustrated in Figs.\,\ref{fig:confined}, \ref{fig:gluonrp}, has a marked influence, e.g., on the $Q^2$-dependence of nucleon-resonance electrocouplings, the extraction of which, via meson electroproduction off protons, is an important part of the current CLAS program and studies planned with the CLAS12 detector \cite{GotheI,MokeevI}.
In combination with well-constrained QCD-based theory, such data can potentially therefore be used to chart the evolution of the mass function on $0.3 \lsim p \lsim 1.2$, which is a domain that bridges the gap between nonperturbative and perturbative QCD.  This might assist in unfolding the relationship between confinement and DCSB.

The appearance of running masses for gluons and quarks is a quantum field theoretic effect, unrealizable in quantum mechanics.  It entails, moreover, that quarks are not Dirac particles and the coupling between quarks and gluons involves structures that cannot be computed in perturbation theory.  Recent progress with the two-body problem in quantum field theory \cite{Chang:2009zb} has enabled these facts to be established \cite{Chang:2010hb}.  One may now plausibly argue that theory is in a  position to produce the first reliable symmetry-preserving, Poincar\'e-invariant prediction of the light-quark hadron spectrum \cite{Chang:2011ei}.

\bigskip

\hspace*{-\parindent}\mbox{\textbf{4.~Mesons and Baryons: Unified Treatment}}~~Indeed, owing to the importance of DCSB, full capitalization on the results of the $N^\ast$-program is only possible within such a framework.  It is essential that one be able to correlate the properties of meson and baryon ground- and excited-states within a single, symmetry-preserving framework, where symmetry-preserving means that all relevant Ward-Takahashi identities are satisfied.  This is not to say that constituent-quark-like models are worthless.  They are of continuing value because there is nothing better that is yet providing a bigger picture.  Nevertheless, such models have no connection with quantum field theory and therefore not with QCD; and they are not ``symmetry-preserving'' and hence cannot veraciously connect meson and baryon properties.

An alternative is being pursued within quantum field theory via the Faddeev equation.  This analogue of the Bethe-Salpeter equation sums all possible interactions that can occur between three dressed-quarks.  A tractable equation \cite{Cahill:1988dx} is founded on the observation that an interaction which describes color-singlet mesons also generates nonpointlike quark-quark (diquark) correlations in the color-antitriplet channel \cite{Cahill:1987qr}.  The dominant correlations for ground state octet and decuplet baryons are scalar ($0^+$) and axial-vector ($1^+$) diquarks because, e.g., the associated mass-scales are smaller than the baryons' masses and their parity matches that of these baryons.  On the other hand, pseudoscalar ($0^-$) and vector ($1^-$) diquarks dominate in the parity-partners of those ground states \cite{Roberts:2011cf}.  This approach treats mesons and baryons on the same footing and, in particular, enables the impact of DCSB to be expressed in the prediction of baryon properties.

\begin{figure}[t]
\includegraphics[clip,width=0.90\textwidth]{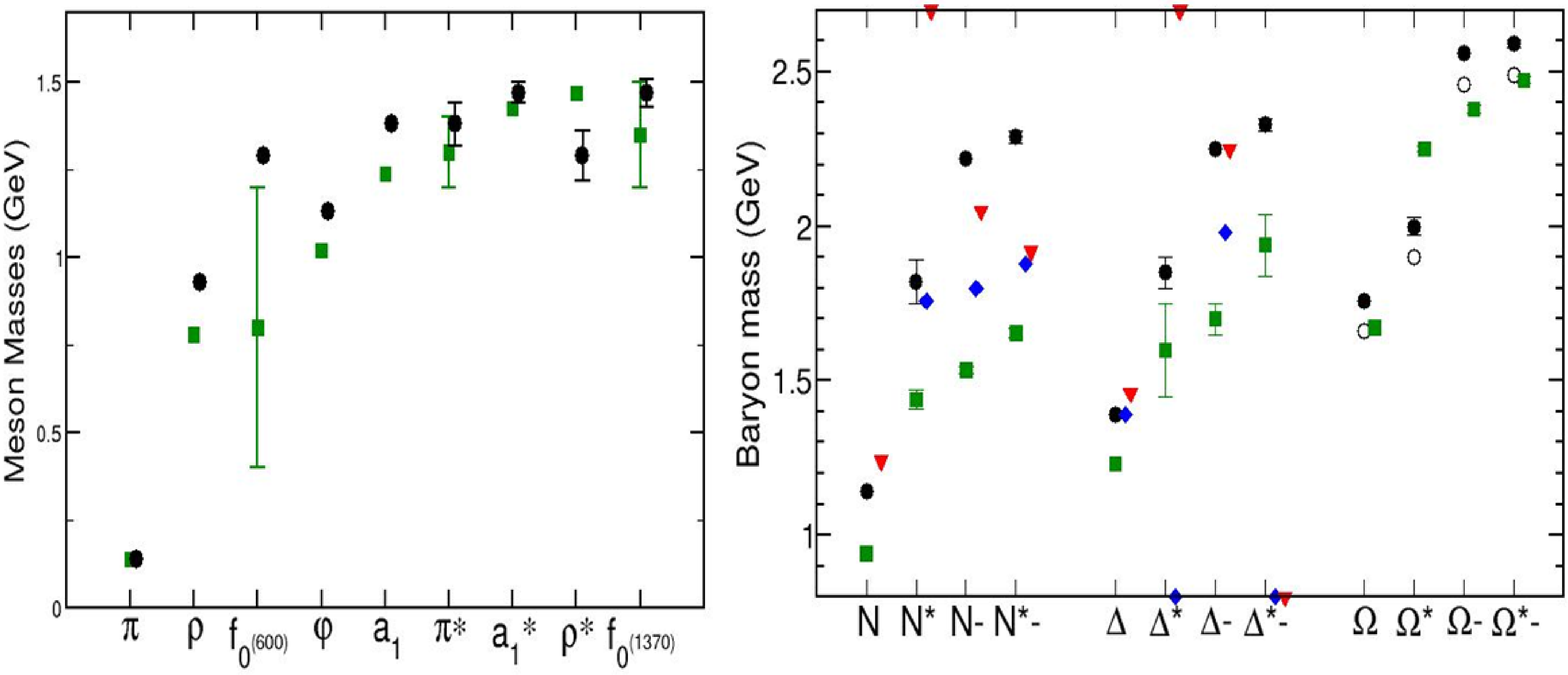}
\caption{\label{Fig2}
Comparison between DSE-computed hadron masses (\emph{filled circles}) and: bare baryon masses from Ref.\,\protect\cite{Suzuki:2009nj} (\emph{filled diamonds}) and Ref.\,\protect\cite{Gasparyan:2003fp} (\emph{filled triangles}); and experiment \protect\cite{Nakamura:2010zzi}, \emph{filled-squares}.
For the coupled-channels models a symbol at the lower extremity indicates that no associated state is found in the analysis, whilst a symbol at the upper extremity indicates that the analysis reports a dynamically-generated resonance with no corresponding bare-baryon state.
In connection with $\Omega$-baryons the \emph{open-circles} represent a shift downward in the computed results by $100\,$MeV.  This is an estimate of the effect produced by pseudoscalar-meson loop corrections in $\Delta$-like systems at a $s$-quark current-mass.
}
\end{figure}

Building on lessons from meson studies \cite{Chang:2011vu}, a unified spectrum of $u,d$-quark hadrons has been obtained using a symmetry-preserving regularization of a vector$\,\times\,$vector contact interaction \cite{Roberts:2011cf}.  This study simultaneously correlates the masses of meson and baryon ground- and excited-states within a single framework.  In comparison with relevant quantities, the computation produces $\overline{\mbox{rms}}$=13\%, where $\overline{\mbox{rms}}$ is the root-mean-square-relative-error$/$degree-of freedom.  As evident in Fig.\,\ref{Fig2}, the prediction uniformly overestimates the PDG values of meson and baryon masses \cite{Nakamura:2010zzi}.  Given that the employed truncation deliberately omitted meson-cloud effects in the Faddeev kernel, this is a good outcome, since inclusion of such contributions acts to reduce the computed masses.

\begin{figure}[t]
\includegraphics[clip,width=0.66\textwidth]{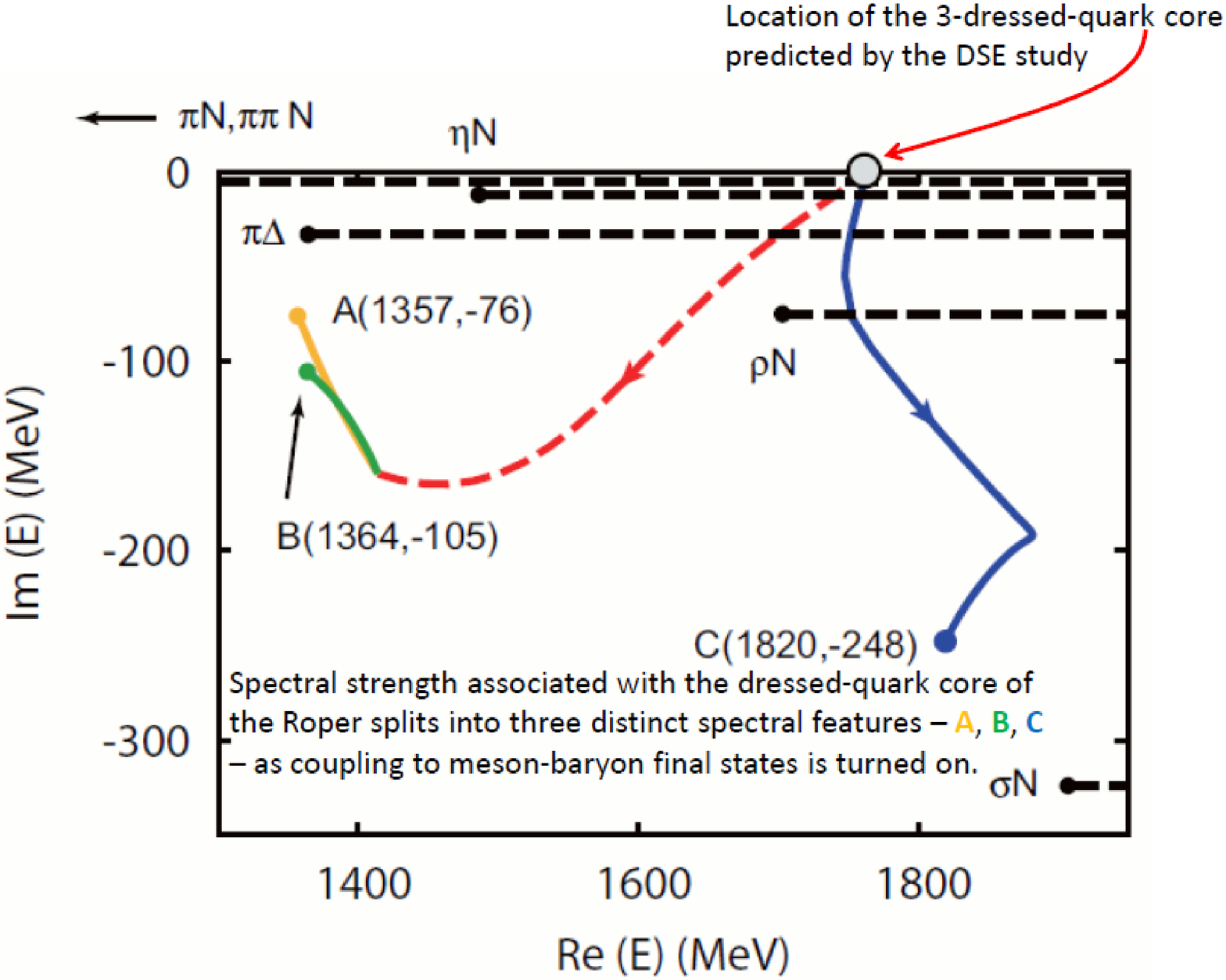}\hspace*{0em}



\parbox{0.3\textwidth}{$\,$\vspace*{-39ex}\\
{\footnotesize \textbf{\footnotesize FIGURE 6}.~~EBAC examined the $P_{11}$-channel and found that the two poles associated with the Roper resonance and the next higher resonance were all associated with the same seed dressed-quark state.  Coupling to the continuum of meson-baryon final states induces multiple observed resonances from the same bare state.  In EBAC's analysis, all PDG-identified resonances were found to consist of a core state plus meson-baryon components.  (Adapted from Ref.\,\protect\cite{Suzuki:2009nj}.)}}

\end{figure}
\addtocounter{figure}{1}

\bigskip

\hspace*{-\parindent}\mbox{\textbf{5.~QCD and Reaction Theory}}~~Following this line of reasoning, a striking result is agreement between the DSE-computed baryon masses \cite{Roberts:2011cf} and the bare masses employed in modern coupled-channels models of pion-nucleon reactions \cite{Suzuki:2009nj,Gasparyan:2003fp}, see Fig.\,\ref{Fig2} and also Ref.\,\cite{CDRobertsII}.  The Roper resonance is very interesting.  The DSE study \cite{Roberts:2011cf} produces a radial excitation of the nucleon at $1.82\pm0.07\,$GeV.  This state is predominantly a radial excitation of the quark-diquark system, with both the scalar- and axial-vector diquark correlations in their ground state.  Its predicted mass lies precisely at the value determined in the analysis of Ref.\,\cite{Suzuki:2009nj}.  This is significant because for almost 50 years the ``Roper resonance'' has defied understanding.  Discovered in 1963, it appears to be an exact copy of the proton except that its mass is 50\% greater.  The mass was the problem: hitherto it could not be explained by any symmetry-preserving QCD-based tool.  That has now changed.  Combined, see Fig.\,6, Refs.\,\cite{Roberts:2011cf,Suzuki:2009nj} demonstrate that the Roper resonance is indeed the proton's first radial excitation, and that its mass is far lighter than normal for such an excitation because the Roper obscures its dressed-quark-core with a dense cloud of pions and other mesons.  Such feedback between QCD-based theory and reaction models is critical now and for the foreseeable future, especially since analyses of CLAS data on nucleon-resonance electrocouplings suggest strongly that this structure is typical; i.e., most low-lying $N^\ast$-states can best be understood as an internal quark-core dressed additionally by a meson cloud \cite{MokeevI}.

The Excited Baryon Analysis Center (EBAC) has made big progress since its inception in March 2006.  It was charged with: completing the combined analysis of available data on single $\pi$, $\eta$ and $K$ photoproduction of nucleon resonances; and incorporation of the analysis of $\pi\pi$ final states.  This will be completed by Spring 2012.  In addition, by the end of 2013 the EBAC collaboration expects to finalize and publish a complete set of well-constrained reaction theory codes.

\bigskip

\hspace*{-\parindent}\mbox{\textbf{6.~Epilogue}}~~This presentation emphasized an international theory effort that works in support of the $N^\ast$-program, now and in connection with the 12\,GeV plans.  The goal is to understand how the interactions between dressed-quarks and -gluons create nucleon ground- and excited-states, and how these interactions emerge from QCD.  This workshop showed that no single approach is yet able to provide a unified description of all $N^\ast$ phenomena; and that intelligent reaction theory will long be necessary as a bridge between experiment and QCD-based theory.  Nonetheless, material progress has been achieved since Beijing~2009: in developing strategies; methods; and approaches to the physics of nucleon resonances.



\bigskip

\hspace*{-\parindent}\mbox{\textbf{Acknowledgments}}~~I acknowledge valuable discussions with V.~Mokeev and S.\,M.~Schmidt, and financial support from the Workshop.
Work supported by:
Forschungszentrum J\"ulich GmbH;
and U.\,S.\ Department of Energy, Office of Nuclear Physics, contract no.~DE-AC02-06CH11357.

\vspace*{-2ex}



\bibliographystyle{aipproc}   


\IfFileExists{\jobname.bbl}{}
 {\typeout{}
  \typeout{******************************************}
  \typeout{** Please run "bibtex \jobname" to optain}
  \typeout{** the bibliography and then re-run LaTeX}
  \typeout{** twice to fix the references!}
  \typeout{******************************************}
  \typeout{}
 }


\end{document}